# Modelling Group Opinion Shift to Extreme :
# the Smooth Bounded Confidence Model

Guillaume Deffuant*, Frédéric Amblard*, Gérard Weisbuch**,

*Cemagref-LISC
24, avenue des Landais
63172 Aubière
France
**ENS-LPS

Contact : guillaume.deffuant@cemagref.fr

*Abstract : We consider the phenomenon of opinion shift to the extreme reported in the social psychology literature. We argue that a good candidate to model this phenomenon can be a new variant of the bounded confidence (BC) model, the smooth BC model which we propose in this paper. This model considers individuals with a continuous opinion and an uncertainty. Individuals interact by random pairs, and attract each other's opinion proportionally to a Gaussian function of the distance between their opinions. We first show that this model presents a shift to the extreme when we introduce extremists (very convinced individuals with extreme opinions) in the population, even if there is the same number of extremists located at each extreme. This behaviour is similar to the one already identified with other versions of BC model. Then we propose a modification of the smooth BC model to account for the social psychology data and theories related to this phenomenon. The modification is based on the hypothesis of perspective taking (empathy) in the context of consensus seeking.*

## 1  Introduction

In 1895, G. Le Bon claimed that crowds tend to adopt much more radical opinions than each individuals would have expressed alone (Le Bon 1895). The famous night of August $4^{th}$, 1789 in the French revolution, where an assembly of nobles decided to abolish their own privileges, was for him a paradigmatic example of this phenomenon. Certainly most individuals participating to this meeting would have never dared to take such radical decisions, had they been alone.

A tremendous set of experiments in social psychology, beginning in the sixties, confirmed the reality of the phenomenon identified by Le Bon (Moscovici and Doise, 1992). When a group is asked to obtain a consensus about a general question (typically about the ability of a political figure or the level of acceptable risk for an enterprise) through a free discussion, the obtained consensus is generally significantly more extreme than the average of the initial individual opinions. This result has been confirmed by hundreds of experiments, performed in a large variety of conditions.

In particular, it was shown that the shift to the extreme is lower when some constraints are imposed to the discussion (such as a discussion protocol, or a time limit to get the consensus). Even if the room where the discussion takes place gives a formal atmosphere (very large room, tables put as in a classroom), the consensus tends to be close to the initial average position, or even more moderate (Forgas 1977, 1981). Moscovici and Doise distinguish therefore between warm discussions (free, informal) leading to extreme consensus, and cold discussions (constrained, formal), leading to moderate, averaging consensus.

Several theories from social psychology were in competition to explain these results (Pruitt, 1971), and experiments were designed in order to get data confirming or falsifying these theories. These theories are so numerous that a complete review is impossible here. We rapidly present in the paper leadership, social comparison and commitment theories, which we relate to our modelling approach.

This rich ensemble of experimental data and theories is a very interesting challenge for individual based modelling approaches. However, rather few researchers from the social simulation communities tried to address this challenge. The most noticeable exception is the work from S. Galam and S. Moscovici who proposed first models of these phenomena, based on variations on the Ising model (Galam & Moscovici, 1991). Despite the high interest of this attempt, we argue that the choice of binary states of opinion is a serious weakness, and that continuous opinion models are more appropriate to account for shift to extreme phenomena.

Among continuous opinion models, the Relative Agreement (RA) model that we proposed recently exhibits unexpected shifts to the extremes (Deffuant et al. 2002), which makes it a good candidate although it was not designed in relation to the mentioned social psychology experiments and theories. In this paper, we propose a new model of social influence, called the "smooth bounded confidence model" which builds on the idea of the bounded confidence (BC) model (Hegselman & Krause, 2002; Deffuant et al., 2000; Weisbuch et al. 2002). Moreover, like the RA model, the smooth BC model has a continuous function of interaction.

All these models consider agents having a continuous opinions (although discrete versions were proposed in (Stauffer et al. 2004)), and an uncertainty. In the initial BC model, individuals interact by random pairs, and attract each other's opinion if the opinion of their interlocutor is at a distance which is lower than the uncertainty of the agent. The smooth BC builds on this idea, and uses an attraction which is proportional to a Gaussian function of the distance between the opinions, and of standard deviation the uncertainty of the agent. We show that this model exhibits the same types of convergence as the RA model, when extremists are introduced into the population at the beginning of the simulations. Moreover, the smooth BC model presents more easily interpretable transitions between the different types of convergence than the RA model.

We discuss the ability of the smooth BC model to fit social psychology experiments and theories on the shift to extreme. Although some arguments play in its favour, we identify some difficulties, in particular the fact that the consensus is not obtained in a significant part of the parameter space.

Therefore, we modify the initial dynamics to take into account the fact the individuals aim at reaching a consensus : we suppose that each individual considers the other not only from his own point of view, but also considers himself from the other's point of view, simulating empathy. Moreover, we suppose that the empathy is asymmetric and depends on the orientation of the axis. For instance individuals empathise more with higher opinion individuals than with lower opinion individuals. We show that this model fits well the social comparison and commitment theories.

In the next section, we summarise some of the experimental data and theories, about the shift to extreme phenomenon. In the third section, we present the smooth bounded confidence model, and study the influence of extremists on a moderate population, on the line of (Deffuant et al. 2002), and discuss its relevance for modelling the shift to extreme phenomenon. In the fourth section, we propose an adaptation of this model (with asymmetric empathy), and we study its behaviour in the social comparison and commitment theory perspectives. The fifth section is devoted to a discussion and some concluding remarks.

## 2  The shift to the extreme phenomenon

### 2.1  Brief summary of experimental data and social psychology theories

Moscovici and Doise (1992) describe a large set of psycho-sociological experiments in which they observe a "radicalisation" or shift to the extreme of opinions as an effect of group discussions. Radicalisation means here that the initial individual opinions (asked to the participants before the discussion), become more extreme in the consensus. This effect is observed when the participants are asked to reach a consensus after the discussion, and if the discussion is free. The experiment was reproduced in many various conditions since the first ones in the sixties. As Moscovici and Doise point out, it is rather surprising, because one would expect a consensus to be a compromise between the initial individual opinions, i.e. an average of the initial opinions. This expectation is widely contradicted by the observations. However, other experiments showed that an averaging consensus takes place when the subjects must comply to a constrained protocol of discussion (hierarchical for instance).

The opinion radicalisation is therefore very sensitive to the context of the discussion. This led to distinguish between "warm" and "cold" discussions. Warm discussions favour free expression of opinions and confrontation, and generally lead to new and radical group consensus whereas cold groups regulate the expression and tend to lead to trade-off average group consensus (Forgas 1977, 1981). Moscovici and Doise underline that it is not possible to give any absolute value judgement about the one or the other process. Radicalisation can lead to violent group behaviours, but also to more imaginative and creative solutions to a problem. Average solution are in general more cautious but also more stereotyped and can lead to big mistakes because of a too strong pressure to conformity. The fascinating result of these experiments is that the orientation of the final consensus can be strongly controlled by the discussion procedure.

Social psychologists propose a large set of concurrent theories to explain this phenomenon (Pruitt, 1971). An important subset of them is called the "value theories", and they have in common the fact that groups shift in a direction toward which most members are already attracted as individuals. They are called "value theories" because they seek the cause of this attraction, and hence the energy behind the shift, in widely held human values. For instance, the social comparison theory assumes that facing a decision, an individual tries to figure out where the other people stand, and wants to be at least at the level of the others in the attractive direction. The commitment theory, supported by Moscovici and Doise, relates the shift to a high commitment to the discussion, in order to explain the distinction between cold and warm discussions. However, it does not provide a precise dynamical process which could be implemented. Our feeling from a first study of this literature is that there is no real consensus on an theoretical explanation of the data.

These social psychology observations and theories offer therefore a very interesting challenge to the social modelling community. What model, based on which theory could fit best the radicalisation in warm groups and the trade-off in cold ones ? Could this model suggest new directions of experiments to social-psychology, or suggest new theories ?

## 2.2   The pioneer modelling attempt by Galam and Moscovici

Galam and Moscovici made a first attempt in this direction. They collaborated in a series of papers (Galam and Moscovici, 1991), in which they consider the Ising model to describe the interactions between participants of a discussion. In their model, each participant has a binary state $s_i$ which is updated according to the state of the others and of a personal as well as a global bias. The standard behavior of these models, especially when the participants are totally connected (every one is related to all the others), is to converge toward a consensus on one or the other binary state ($+1$ or $-1$). However, with some distributions of individual biases, some participants may change of state less easily, or even always remain in their initial state. When such individuals are distributed in the population, the final result of the discussions can be heterogeneous: some participants have the state 1 and the others the state $-1$.

Galam and Moscovici interpret the final consensus state of the group as the average of the individual states. With such an interpretation, the model can lead to radical consensus (when every body agrees on one or the other binary state), or trade-off consensus, when different states are found in the population after the convergence. In this perspective, the Ising model can be used to model the experiments of psycho-sociology.

However, in our view, using binary state individuals to model these social phenomena is not very appropriate. In the psycho-sociological experiments, the individuals are asked an opinion which can have 15 modalities (from very against to very for). The change of individual opinions after the discussions is measured on this scale. The variable playing the role of this individual opinion is not obvious in the Ising model. If the opinion of an individual is considered to be the average of all states in the population at the end of the simulations (as proposed by Galam), then it should be the same in the previous time steps. But in this case, the population is always in a consensus, which does not fit the data.

It seems therefore interesting to consider continuous opinion models, and explore how they could model the considered experiments.

# 3 The smooth Bounded Confidence model

## 3.1 The interaction rules

The bounded confidence (BC) model considers agents having a continuous opinion and an uncertainty (Deffuant et al. 2001, Hegselmann and Krause, 2002). The principle of these models is that an agent takes into account opinions from others in a limited zone (defined by the uncertainty), around its own opinion. Stauffer et al. (Stauffer et al., 2004) proposed a version of this model with discrete modalities, showing very similar dynamical properties. In (Deffuant et al., 2002) we criticised the BC dynamics, because the influence function is not continuous, and is maximum when the influencing opinion is at the limit of the uncertainty segment. This feature seems psychologically rather unrealistic. We proposed an evolution of this opinion dynamics, the relative agreement model (RA model), in which the influence varies continuously when the difference of opinion varies .

We studied the behaviour of the RA model when we introduce extremist agents into the population : these agents have opinions located at the extremes of the opinion axis (+1 or –1), and they have a low uncertainty. The other agents, with opinion distributed uniformly on [-1,+1] have a higher uncertainty, and are called "moderate agents". We observe then three types of convergence, which mainly depend on the initial uncertainty of the moderate agents:

- Central convergence: The moderate agents evolve and form central clusters, being only marginally influenced by the extremists;
- Double extreme convergence: The moderate agents split into two clusters, each one converging to one of the extreme;
- Single extreme convergence: Almost all the moderate agents are attracted by one of the extremes.

We studied this model with the assumption that every pair of agents could interact (Deffuant et al., 2002), and in (Amblard and Deffuant, 2004) we studied the influence of different social network topologies on the model behaviour, especially the possibility for single extreme convergence.

We propose now to study an other variant from the BC influence function, which aims at smoothing the influence function. We suppose that the influence of an other opinion is weighted by a Gaussian function with a standard deviation equal to the uncertainty of the individual. An individual of opinion $x(t)$ and uncertainty $u(t)$ being influenced by an individual of opinion $x'(t)$ and uncertainty $u'(t)$, will have the opinion $x(t+1)$ and uncertainty $u(t+1)$ after the discussion, which is defined in equations 2 and 3.

We first define :

$$g_u(x-x') = \exp\left(-\left(\frac{x-x'}{u}\right)^2\right) \quad (1)$$

This function is used to compute the influence of an encountered opinion $x'(t)$ on the individual's opinion $x(t)$ and uncertainty $u(t)$:

$$x(t+1) = \frac{x(t) + x'(t).g_u(x(t)-x'(t))}{1 + g_u(x(t)-x'(t))} \quad (2)$$

$$u(t+1) = \frac{u(t) + u'(t).g_u(x(t)-x'(t))}{1 + g_u(x(t)-x'(t))} \quad (3)$$

In this model, the function of influence is smooth when $x'$ and $u'$ vary. Moreover, small uncertainty opinions are more influential than high uncertainty opinions. This model shows the characteristics we were looking for by proposing the RA model, with a more direct mathematical expression.

## 3.2 Convergence

As expected, its dynamical properties are the similar to the ones of the RA model with the presence of extremists in the population. Figures 1, 2 and 3 show examples of the 3 convergence types, obtained with this model.

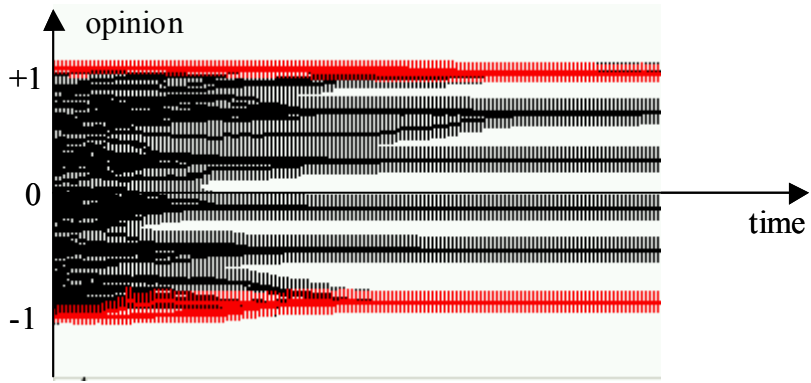

Figure 1 : Central convergence of Smooth BC model with uncertainty of moderate individuals : 0.1, of extremists (represented in red) 0.05, and number of extremists : 5 on each extreme, for 100 moderate. In this case the extremists attract only a small part of the moderate (indicator $y = 0.08$). The vertical segments represent the agent uncertainties.

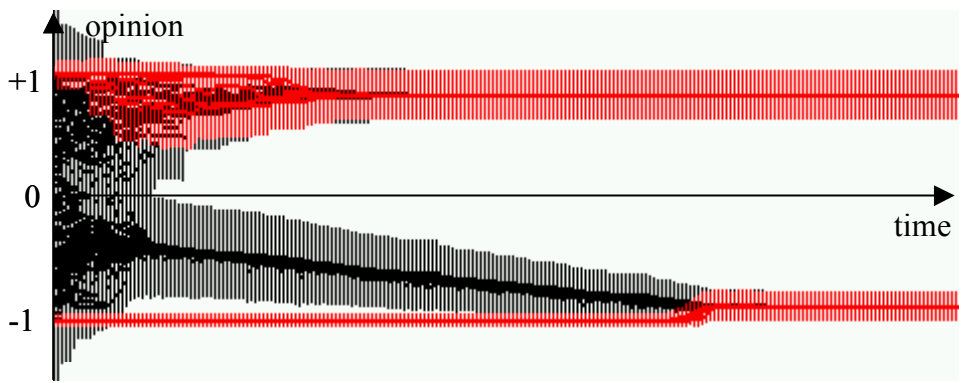

Figure 2 : Double extreme convergence of smooth BC model with uncertainty of moderate individuals : 0.5, of extremists (represented in red) 0.05, and number of extremists : 5 on each extreme, for 100 moderate. In this case each extreme attracts about half of the moderate (indicator $y = 0.53$). The vertical segments represent the agent uncertainties.

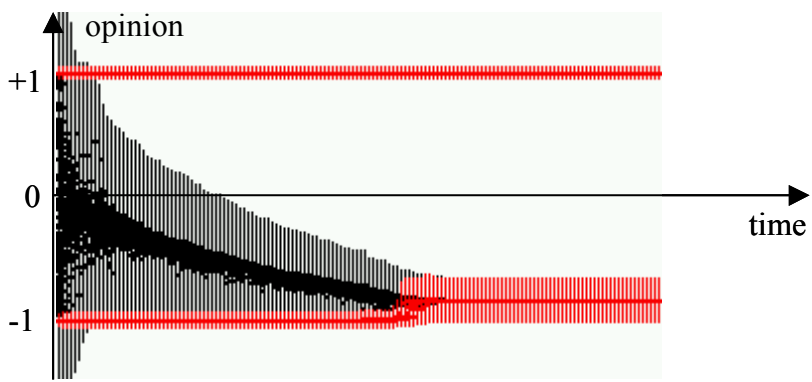

Figure 3 : Single extreme convergence of smooth BC model with uncertainty of moderate individuals : 1.2, of extremists (represented in red) 0.05, and number of extremists : 5 on each extreme, for 100 moderate. In this case the extremists located at –1 attract all the moderate (indicator $y = 0.83$). The vertical segments represent the agent uncertainties.

We define the indicator $y$ of the convergence type as a variant of the one proposed in (Deffuant et al. 2002). Considering a final state of the population involving $k$ clusters of opinion $x_i$, including a proportion $p_i$ of the initially moderate, the indicator is defined by:

$$y = \sum_{i=1}^{k} p_i^2 \cdot |x_i| \quad (4)$$

This indicator is designed in such a way that $y$ is close to 0 when the moderate are not attracted by the extremes, close to 0.5 for double extreme convergence and close to 1 for single extreme convergence. Plotting indicator $y$ when $U$ the uncertainty of the moderate and $p_e$ the proportion of extremists vary yields the graph of figure 4.

The main features of Figure 4 are similar to the one obtained with the Relative Agreement model. The single extreme convergence ($y$ close to 1) takes place for high uncertainty of the moderate, and relatively small initial proportion of extremists. The double extreme convergence takes place for medium of high uncertainty of the moderate, but in the latter case, the proportion of extremists should be high. The central convergence takes place when the uncertainty of the moderate is small.

However, one can also notice a significant difference : with the RA model, the parameter zone between double and single extreme convergences shows a high probability of central convergence, for which it is difficult to find a psychological or social interpretation. The picture with the smooth BC does not exhibit such a zone, and this is an argument in favour of the smooth BC model.

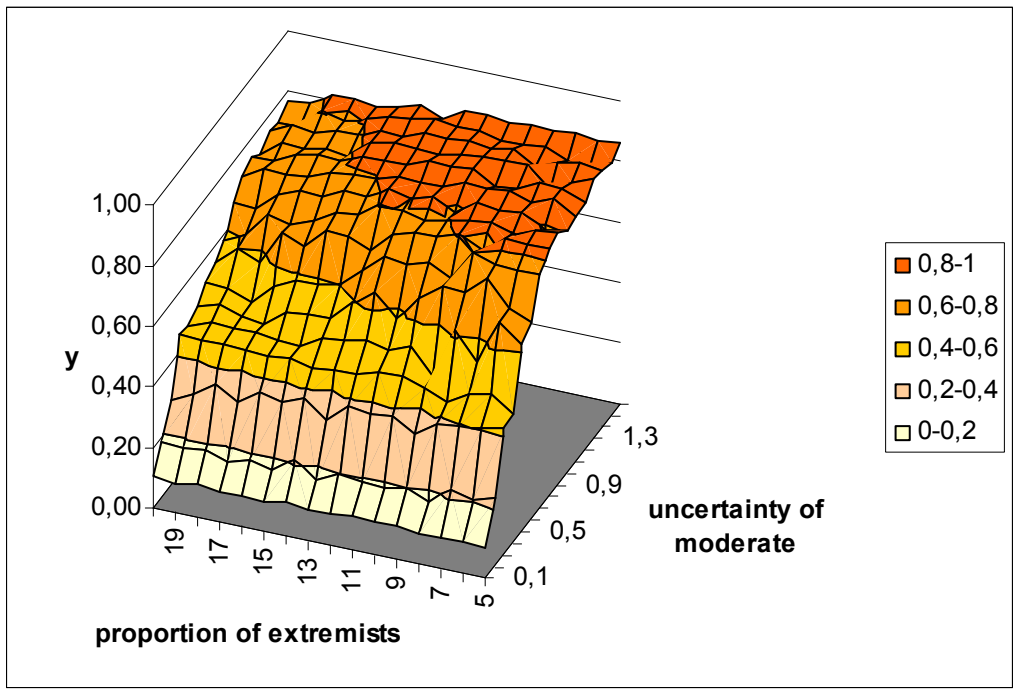

Figure 4 : Convergence indicator for smooth BC, when the proportion of extremists and the initial uncertainty of the moderate vary. The result is the average on 20 replicas. The single extreme convergence (shift to the extreme, $y$ close to 1) appears when the uncertainty of the moderate is high, and when the proportion of extremists is not too high.

The most surprising result of this study, the single extreme convergence which takes place with the same number of extremists at each extreme, can be seen as a particular case of shift to extreme, although this model with extremists was designed independently from the shift to extreme phenomenon studied in social psychology. To determine whether this apparent similarity goes beyond a superficial analogy requires a closer analysis.

# 4 Using the smooth BC to model the shift to extreme phenomenon

## 4.1 The leadership theory

In the set of leadership theories (several variants were proposed, see Pruitt 1971 for a review), the shift to the extreme is due to a higher certainty of the individuals with extreme opinions. These individuals tend to attract the moderate individuals, who are less certain. Some empirical evidence support this theory. It was observed for instance that the most extreme individuals are generally more certain than the others and if they are not, the shift to the extreme does not take place.

The smooth BC (and RA) model with extremists, are particularly close to this theory. The continuous opinion gives a direct interpretation to more or less extreme opinions, and the uncertainty variable finds a natural interpretation, as a measure of conviction (ability to convince and to be convinced). Considering extremists as very convinced agents, with extreme opinions fits very closely the leadership theory.

Moreover, the smooth BC model can easily distinguish between the radicalisation in warm discussions and the trade-off in cold ones. Suppose that the groups include a small set of more convinced individuals with nearly extreme opinion, because of a fundamental tendency of the values associated to this opinion axis, and the others are moderate, then the behaviour of the RA model is:

- convergence to one extreme (radicalisation) if the extremists are very convinced and the moderate very uncertain (figure 4).

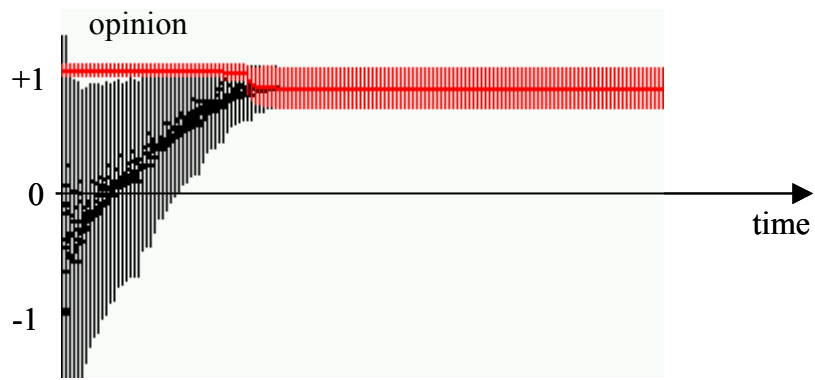

Figure 4 : Shift to the extreme with smooth BC, with extremists only on one side. 10 moderate individuals with initial uncertainty of 1.2 are attracted by a single leader with initial uncertainty 0.05. The extreme opinion individual attracts all the moderate ones.

- central convergence (trade-off) if the difference of uncertainty between extremists and moderate is not too high, and all are uncertain (figure 5).

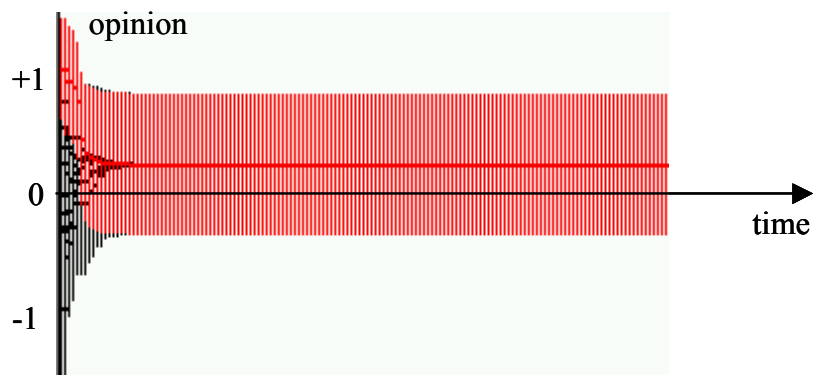

Figure 5 : Averaging consensus obtained with an extreme individual with uncertainty 0.4, and 10 moderate ones with uncertainty 0.6. The final consensus is closed to the average of initial opinions.

- A set of opinion clusters if all are certain of their opinions.

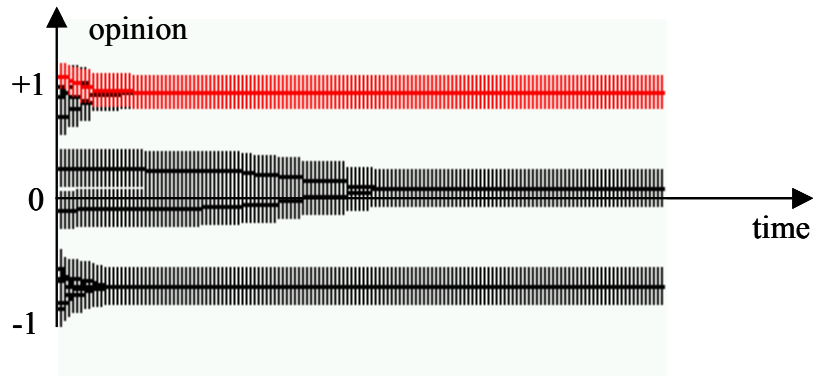

Figure 6 : Set of opinion clusters obtained with an extreme individual with uncertainty 0.1, and 10 moderate ones with uncertainty 0.15. In this case, there is no final consensus.

A possibility of interpretation is therefore that :

- in warm groups the individuals more freely express their certainties or uncertainties which results in a perceived higher difference between extremists and moderates. This difference leads to a radicalisation in the smooth BC model.
- On the contrary, in cold groups the expression of certainty or uncertainties is inhibited which leads to trade-off consensus.

The interpretation of the difference between warm and cold groups seems reasonable in the smooth BC and RA perspective. One can expect that a formal protocol of discussions will limit the expression of opinions, and of very large uncertainties as well as very strong certainties, which on the contrary will freely be expressed in an unconstrained context.

However, this approach is not totally satisfactory because the commitment to reach a consensus is not taken into account into the model, and we can easily get situations such as in figure 6 where the consensus is not reached. Moreover, it seems to contradict the commitment theory.

## 4.2 Commitment and social comparison theories

The commitment theory, advocated by Moscovici and Doise, shows some difference with the leadership theory. It relies on the fact that during warm discussions, individuals get more and more involved, and this commitment follows the existing asymmetry of the considered value axis, resulting in a shift to the extreme. The point is that all the individuals tend to get more involved and more convinced, during warm discussions. On the contrary, during cold discussions, individuals are not much involved, and all of them tend to be less convinced. Therefore, Moscovici and Doise, in their commitment theory, claim that the shift to the extreme is due to a higher involvement of all the participants. The precise mechanism leading to radicalisation or averaging is not precisely described.

If we consider that the uncertainty also represent the inverse of the conviction or commitment, then the smooth BC is in contradiction with the commitment theory. In effect, the commitment theory would predict a strong shift to the extreme in the condition of figure 6, because all individuals are very convinced and committed.

Although the commitment theory mechanism is not described in details, it can be related more easily to value theories or social comparison theories. The "social comparison" is at the basis of a whole set of social psychology theories aiming at explaining the shift to the extreme phenomenon. These theories suppose that it is somewhat uncomfortable to be seen below the others on the axis of opinion (or above depending on the value orientation of the axis). This leads individuals to increase their opinions when they observe that they are below the others, and induces progressively a shift to the attractive extreme. It is interesting to see whether our model can adapted to this theory.

## 4.3 Social comparison and smooth BC with asymmetric empathy

To take into account these theories, we propose to modify the smooth BC dynamics by supposing that each individual not only considers the opinion of the other from his own point of view, but also takes into account the point of view of the other on his own opinion. The equations giving $x(t+1)$ as a function of $x(t)$, $x'(t)$, $u(t)$ and $u'(t)$ include now a parameter $\mu$ which weights the influence of each point of view as specified in equations 5 and 6 ($g_u$ is the Gaussian function defined by equation 1):

$$x(t+1) = (1-\mu)\frac{x(t)+x'(t).g_u(x(t)-x'(t))}{1+g_u(x(t)-x'(t))} + \mu\frac{x'(t)+x(t).g_{u'}(x(t)-x'(t))}{1+g_{u'}(x(t)-x'(t))} \quad (5)$$

$$u(t+1) = (1-\mu)\frac{u(t)+u'(t).g_u(x(t)-x'(t))}{1+g_u(x(t)-x'(t))} + \mu\frac{u'(t)+u(t).g_{u'}(x(t)-x'(t))}{1+g_{u'}(x(t)-x'(t))} \quad (6)$$

We model the social comparison by introducing an asymmetry in the empathy : an individual has empathy for opinions which are above his own which are weighted by $\mu$, and for opinions which are below his own are weighted by $\alpha.\mu$ (with $\alpha < 1$). This implies that empathy is higher for higher opinions than for lower opinions (of course this can be reversed if the value axis is oriented in the opposite side), as illustrated by figure 6.

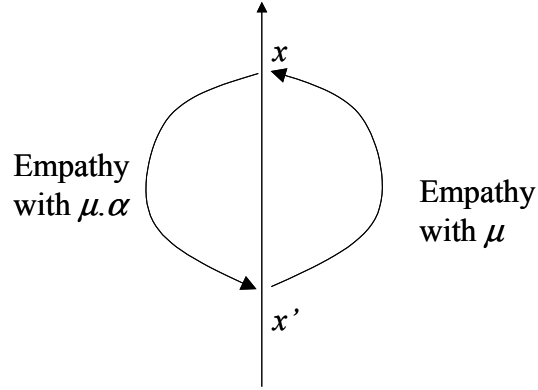

Figure 6 : Asymmetry of empathy. Individuals have less empathy ($\alpha < 1$) for individuals with opinions below theirs, than for individual with opinions above theirs.

Moreover, instead of considering pair interactions, we suppose that each individual talks to all the others (however this does not change much the final result). Figure 7 shows an example of evolution of the smooth model with asymmetric empathy, with individuals having small uncertainties. We observe significant shifts to the extreme.

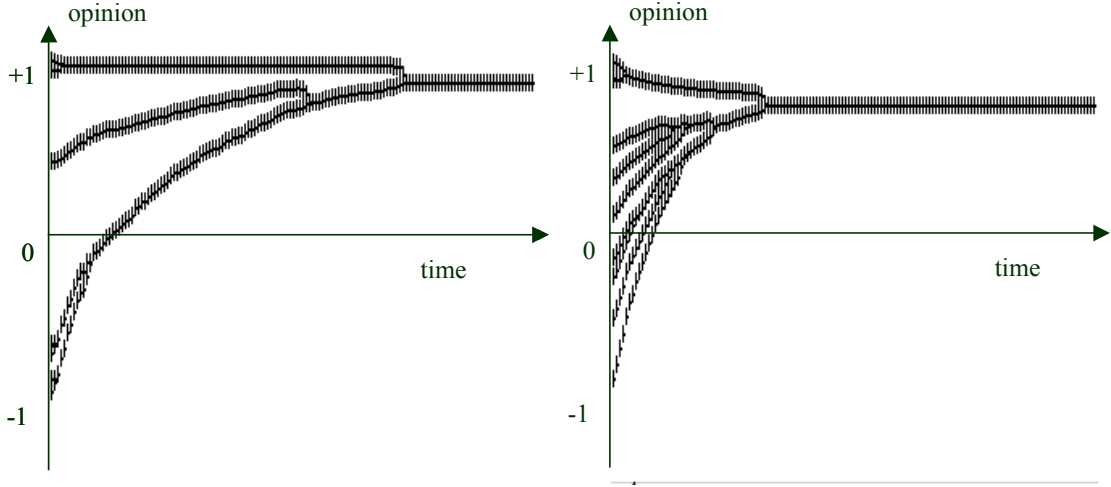

Figure 7 : smooth BC with asymmetric empathy. 10 individuals with uncertainty 0.05. $\mu = 0.01$. On the left $\alpha = 0$, on the right, $\alpha = 0.1$. We observe significant shifts to the extreme, whereas the uncertainties are low.

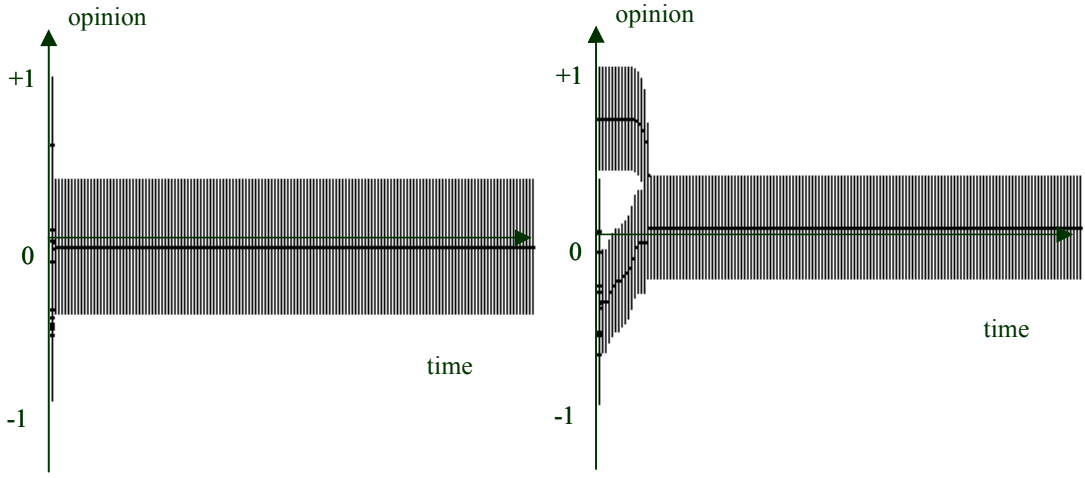

Figure 8 : Smooth BC with asymmetric empathy, and larger uncertainties. 10 individuals, $\mu = 0.01$, $\alpha = 0$. On the left the uncertainty of individuals $U$ is 0.4, on the right, 0.3.

On figure 9, we plotted a systematic exploration of the shift to extreme when $\alpha$ and $U$ vary. The shift is measured as follows : Considering $N$ individuals of initial opinions $x_i(0)$ and of final opinions $x_i(T)$, for $i$ between 1 and $N$, the shift to the extreme $s$ is computed by the formula of equation 7.

$$s = \frac{2\left|\sum_{i=1}^{N} x_i(T) - \sum_{i=1}^{N} x_i(0)\right|}{\max_i(x_i(0)) - \min_i(x_i(0))} \qquad (7)$$

We note that the important shifts take place only when both $\alpha$ and $U$ are small, and that we find smaller shifts than in presence of an extremist, without asymmetry of empathy. Significant shifts to the extreme take place only when $\alpha$ is small, that is when the value axis is strongly oriented.

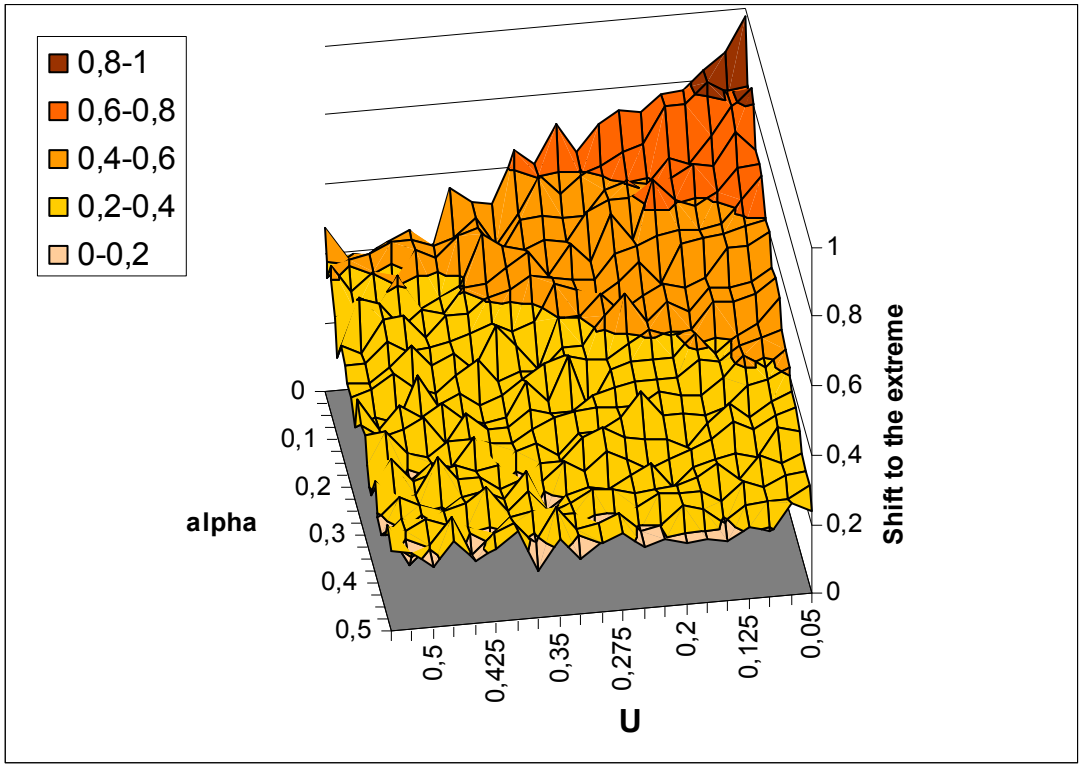

Figure 9 : Shift to the extreme (see equation 7) for smooth BC with asymmetric empathy. $\alpha$, representing the difficulty of empathy for lower opinion and $U$ the uncertainty of the individuals vary. The shift is maximum when $U$ and $\alpha$ are small.

## 5  Conclusion

Our work is initially motivated by the behaviour of the Relative Agreement (RA) model with extremists : this model shows a shift to the extreme when the uncertainty of moderate individuals is large, even when the number of extremists is the same at each extreme. We did not find experiments in the social psychology literature which correspond to the behaviour of this model. However, a huge literature is devoted to a shift to extreme phenomenon, which takes place in condition of free discussion aiming at achieving a consensus.

In this paper, we propose a new opinion influence model, the smooth BC model, which uses a Gaussian function instead of a step function to rule the opinion influence. We show that this model has the same convergence types as the RA model, and that the transitions between the different convergence types are easier to interpret than with the RA model. We proposed a first interpretation of the shift to the extreme which supposed that the individuals with more extreme opinions are more certain.

However, we criticised this interpretation, because it seems to contradict the commitment theory, and it does not take into account the fact that a consensus is required. Therefore, we proposed a new version of the smooth BC model, introducing the other's point of view on oneself as a source of opinion evolution (we call it empathy), which is particularly important when the consensus is required. To take into account the value orientation which is assumed in the social psychology theories, we suppose that this empathy is asymmetric : individuals empathize less with lower opinion individuals than with higher opinion ones.

The smooth BC model with asymmetric empathy shows results which are compatible with the commitment theory : the highest shifts to the extreme appear when the individuals are certain.